\documentclass[12pt,preprint]{aastex}

\usepackage{rotating,wasysym}

\newcommand{\msun}{\ensuremath{{\mathrm{M}_{\odot}~}}}

\shorttitle{Decoding the message from stardust SiC grains}
\shortauthors{Lewis et al.}

\begin{document}


\title{Decoding the message from meteoritic stardust silicon carbide 
grains}


\author{Karen M. Lewis}

\affil{Monash Centre for Astrophysics (MoCA), Monash University,
Clayton VIC 3800, Australia}
\affil{Earth and Planetary Sciences, 
Tokyo Institute of Technology, Japan}
\email{karen.michelle.lewis@gmail.com}

\author{Maria Lugaro}

\affil{Monash Centre for Astrophysics (MoCA), Monash University,
Clayton VIC 3800, Australia}
\email{maria.lugaro@monash.edu}

\author{Brad K. Gibson and Kate Pilkington}

\affil{Jeremiah Horrocks Institute, University 
of Central Lancashire, UK}
\affil{Monash Centre for Astrophysics (MoCA), Monash University,
Clayton VIC 3800, Australia}
\email{bkgibson@uclan.ac.uk, kpilkington@uclan.ac.uk}



\begin{abstract}

Micron-sized stardust grains that originated in ancient stars are 
recovered from meteorites and analysed using high-resolution mass 
spectrometry. The most widely studied type of stardust is silicon carbide 
(SiC). Thousands of these grains have been analysed with high 
precision for their Si isotopic composition. Here we show that the 
distribution of the Si isotopic composition of the vast majority of 
stardust SiC grains carry the imprints of a spread in the age-metallicity 
distribution of their parent stars and of a power-law increase of the 
relative formation efficiency of SiC dust with the metallicity. This result 
offers a solution for  
the long-standing problem of silicon in stardust SiC grains,  
confirms the necessity of coupling chemistry and dynamics in simulations 
of the chemical evolution of our Galaxy, and constrains the modelling of dust 
condensation in stellar winds as function of the metallicity.

\end{abstract}



\section{Introduction} \label{sec:intro}

A small fraction (order of 1-100 parts per million in mass) of 
the matrix of 
primitive meteorites is composed of {\it stardust} grains. These grains 
originated in stars, were present at the formation of the Solar System, 
and have been preserved inside meteorites in their original form until 
today. Since their discovery in the late 1980s stardust grains have been 
extensively analysed and employed to constrain our understanding of 
nucleosynthesis, mixing, and dust formation in stars and supernovae, 
Galactic chemical evolution (GCE), processing of dust in the 
interstellar medium, the formation of the Solar System, and the 
evolution of meteorite parent bodies \citep{redbook,clayton04}. 
A large variety of 
minerals have been discovered as stardust, from diamond, to silicate
and Al$_2$O$_3$ grains. Among them, silicon carbide (SiC) grains 
have been 
the most widely studied both due to their relatively large size (up to 
several $\mu$m), as compared to other types of stardust, and to the 
relatively easier separation procedure. The vast majority of the 
stardust SiC grains recovered from meteorites show the clear signature 
of an origin in the mass-losing envelopes of asymptotic giant branch 
(AGB) stars \citep{gallino90,lugaro03a} that become C rich 
(C$>$O) due to dredge-up of material from the deep C-rich layers into 
the convective envelope of the star. In C-rich conditions some C is free 
from the strong CO molecular bond and can react with Si to form SiC. 
Stardust SiC 
grains of C-rich AGB origin are the ``mainstream'' population, which 
comprises $>$93\% of all stardust SiC, and the minor Y and Z 
populations, which comprise $\simeq$1\% each of all stardust SiC. 
The Si compositions of mainstream, Y, and Z grains carry the signature 
of both the initial composition of their parent star, which is 
determined by GCE, and the neutron-capture and mixing processes that 
occurred in the AGB parent stars \citep{zinner06}. 
The typical Si isotopic ratios of the Y and Z grains point to AGB parent 
stars of average metallicity $\simeq$ 1/2 and $\simeq$ 1/3, 
respectively, of solar, while the Si isotopic ratios of mainstream 
grains suggest a close-to-solar metallicity for their parent 
stars \citep{hoppe97,amari01a,zinner06}.

\begin{figure} 
\includegraphics[width=0.9\textwidth]{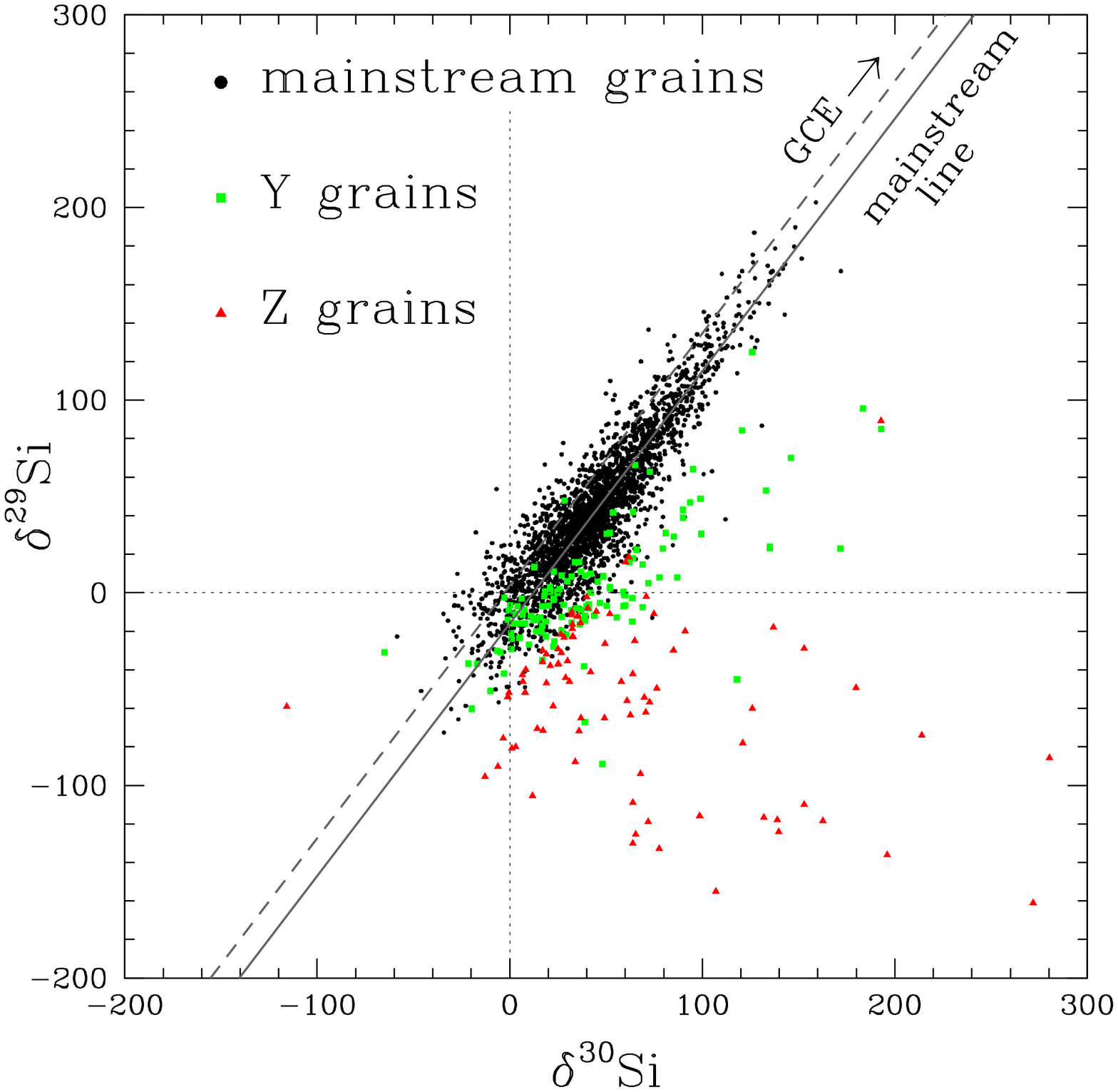}
\caption{Distribution of the Si isotopic compositions of SiC 
mainstream, Y, and Z grains with 1$\sigma$ error bar lower than 
15$\permil$ represented using the $\delta^{29,30}$Si 
notation, i.e., the permil variation of $^{29,30}$Si/$^{28}$Si with respect 
to the 
solar ratios. The solid line is the best fit through the mainstream SiC data 
(``mainstream line'') of slope 1.31 and intercept $-15.9\permil$.
The dashed line is the mainstream line shifted by $-15\permil$ in 
$\delta^{30}$Si. This is taken to represent the GCE of the Si isotopic ratios, 
with $\delta^{29,30}$Si increasing with the metallicity. 
\label{fig:SiCdata}}
\end{figure}

A large amount of high-precision Si isotope data from stardust 
SiC has been collected in 
the past 25 years (Figure~\ref{fig:SiCdata}), however, their 
distribution is not understood. A particularly irksome problem is that 
most mainstream grains have $^{29}$Si/$^{28}$Si and $^{30}$Si/$^{28}$Si 
larger than 
solar (up to +20\%) while according to GCE models the 
$^{29}$Si/$^{28}$Si and $^{30}$Si/$^{28}$Si ratios increase with 
metallicity, which in turn increases with time \citep{timmes96}. 
The grains must have 
formed in stars of metallicity higher than solar, however, their parent 
stars must have died before the Solar System formed. Several possible 
explanations have been proposed for this apparent paradox, from a simple 
model of stellar migration from the inner part of the 
Galaxy \citep{clayton97}, which has difficulties in reproducing the 
observed distribution \citep{nittler99}, to inhomogeneities in the 
interstellar medium \citep{lugaro99}, which is at odds with the 
correlation between the Si and Ti isotopic composition of the 
grains \citep{nittler05}, to a starburst triggered by the merging of our 
Galaxy with another galaxy \citep{clayton03}. To address this problem, 
we use the measured 
$\delta^{29}$Si and $\delta^{30}$Si of SiC grains from AGB stars 
(Figure~\ref{fig:SiCdata}) to derive the age-metallicity relation (AMR) of 
their parent stars and compare it to that observed for stars in the 
solar neighbourhood.

\section{Method} \label{sec:method}

We selected from the Presolar Grain Database \citep{hynes09} the 2,732 
mainstream, 133 Y, and 92 Z grains with 1$\sigma$ error bar lower than 
15$\permil$. For each grain we applied the following steps:

\begin{enumerate}

\item{ 
We inferred the metallicity [Fe/H] (defined as the logarithm of the Fe/H 
ratio with respect to solar) of the parent star of each SiC grain from 
the relationship between $\delta^{29}$Si and [Fe/H] predicted by 
different GCE models (Figure~\ref{fig:GCEmodels}). All the models have 
been renormalised so that at the time of the formation of the Sun, set 
to 8.5 Gyr, [Fe/H]=0 and $\delta^{29}$Si=0, by definition. Before 
normalisation, all the models produce $\delta^{29}$Si
between $-600$ and $-400$ at [Fe/H]=0, a long-standing problem probably
related to
the rates of the nuclear reactions that produce $^{29}$Si in core-collapse
supernovae \citep{hoppe09}.} 

\item{We estimated the change in 
$\delta^{30}$Si resulting from AGB nucleosynthesis as the distance 
$\Delta^{30}$Si between the measured $\delta^{30}$Si and the value 
obtained from the GCE line. We assumed that the best fit to the Si 
isotopic ratios in the Si 
three-isotope plot (the ``mainstream line'') shifted by $-$15 in 
$\delta^{30}$Si represents the GCE of the Si isotopic ratios 
(Figure~\ref{fig:SiCdata}). This shift is in agreement with the Si 
composition of stardust silicate grains from AGB stars, which 
represents the Si composition of O-rich AGB stars as unaltered by 
nucleosynthesis and dredge-up \citep{mostefaoui04,nguyen10}.} 

\item{We derived the parent star age from its mass as obtained from
$\Delta^{30}$Si using the set of FRANEC C-rich AGB models with the
neutron-capture cross sections of the Si isotopes by \citet{guber03}
presented in \citet{zinner06}, to which we added the age of the Sun.
In these models the mass-loss rate was included using the
parametrization given by \citet{reimers75}. For the 1.5 \msun and 2
\msun models results were presented for different values of the
associated free parameter $\eta$=0.1, 0.3, 0.5 and we choose to 
average the results for each mass. 
A range of $\Delta^{30}$Si is allowed during the
C-rich phase of each model since $\Delta^{30}$Si increases with
the number of dredge-up episodes. This results in different
possible masses associated to the same
$\Delta^{30}$Si, in particular when its value is small.
We chose to remove this degeneracy by taking as the best representative
of each model the Si composition reported after the
last computed dredge-up episode. 
Most SiC grains formed with the composition present in the envelope after
the final few dredge-ups since the largest fraction
of the envelope mass during the C-rich phase is lost in these 
final phases. For example,
considering the 3 \msun model with $Z=0.02$, more than 2/3 of the envelope mass
during the C-rich phase is lost after the third-last TDU episode,
and roughly 1/2 is lost after the very last TDU episode \citep[see
Table 4 of][]{straniero97}. Note that we did not include in
our age determination the up to 1 Gyr of grain residence time in the
interstellar medium \citep{gyngard09}.}

\item{Since also $\delta^{29}$Si can 
be marginally affected by AGB nucleosynthesis we improved the estimates 
of age and metallicity by repeating the same procedure as above with a 
new initial $\delta^{29}$Si$_{new}$ = $\delta^{29}$Si - $\Delta^{29}$Si, 
with $\Delta^{29}$Si derived from the same AGB model predictions used to 
match $\Delta^{30}$Si.}

\end{enumerate}

\section{Results and discussion} \label{sec:results}

The resulting SiC AMR is plotted in Figure~\ref{fig:AMR} together with 
the AMR derived for stars in the solar neighbourhood from the 
Geneva-Copenhagen (G-C) survey \citep{holmberg07}. 

The SiC ages are affected by several uncertainties. First, there are 
random uncertainties related to the measurement errors. 
We made the conservative choice to plot
the lower limits of the ages derived from adding the 
experimental 2$\sigma$ error bar 
to $\delta^{30}$Si. 
Second, there are random errors related to the possible 
effect of inhomogeneities in the interstellar medium, which may change the 
Si composition of any given parent stars by $\sim50\permil$ 
\citep{lugaro99,nittler05}. These are not possible to be evaluated. 
Assuming that the silicon isotopic distribution is 
relatively symmetric, the age and the metallicity calculated 
for grains at the peak of the distribution should be reliable.
Third, there are systematic uncertainties related to the choice 
of the line taken 
to represent the GCE in the Si isotope plot (Figure~\ref{fig:SiCdata}) and to 
the AGB model predictions. The uncertainty related to the GCE line 
would most likely result in smaller stellar ages as the line could be 
shifted further away from the mainstream line than what we have 
assumed and still be in agreement with the silicate data. Instead, it 
is not known if the uncertainty related to the AGB models would result 
in smaller or larger stellar ages since both larger and smaller 
$\Delta^{30}$Si for a given stellar mass are possible within the 
uncertainties wrought by, e.g., the mass-loss rate, the efficiency of 
the dredge-up of the deep layers of the stars into the convective 
envelope, and the neutron-capture cross section of the Si isotopes.

\begin{figure}
\includegraphics[width=0.9\textwidth]{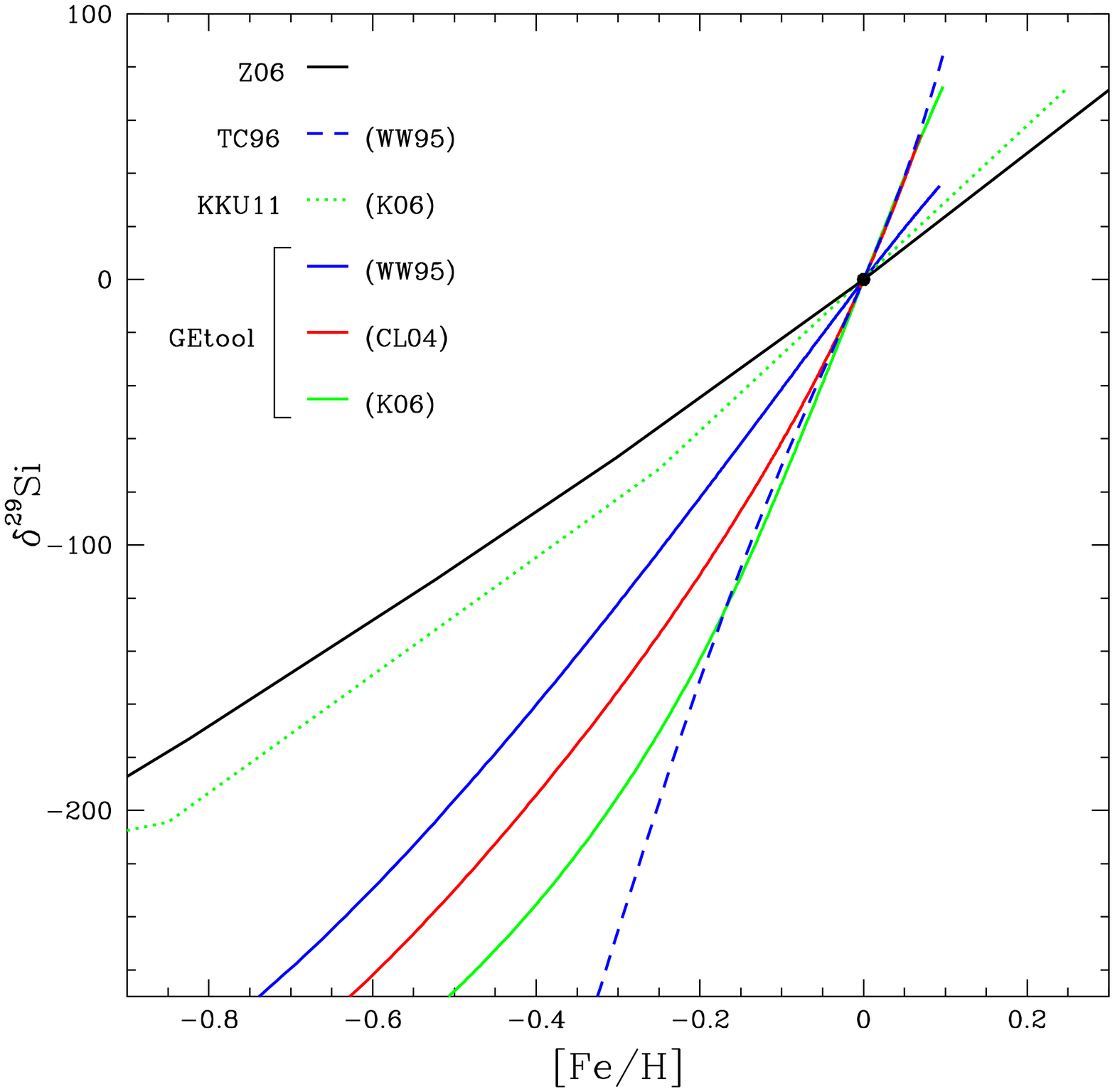}
\caption{
The relationship between $\delta^{29}$Si and [Fe/H] derived from different
GCE models: TC96 \citep{timmes96}, KKU11 \citep{kobayashi11a}, and 
GEtool \citep{fenner03a}, 
using different SNII yields: WW95 \citep{woosley95}, CL04 \citep{chieffi04}, 
and 
K06 \citep{kobayashi06}; or by simply assuming 
that the abundances of 
$^{29}$Si and $^{30}$Si scale with the metallicity, while that of 
$^{28}$Si is $\alpha$ enhanced such that $^{28}$Si is 1/8 of its
solar abundance when the metallicity is 1/10 of solar: Z06 \citep{zinner06}. 
This choice produces a $\delta^{29}$Si and [Fe/H] relationship close to that 
obtained by \citet{zinner06} comparing the composition of Z grains to AGB 
models 
of fixed mass and variable metallicity. 
\label{fig:GCEmodels}}
\end{figure}

\begin{figure}
\includegraphics[angle=270,width=\textwidth]{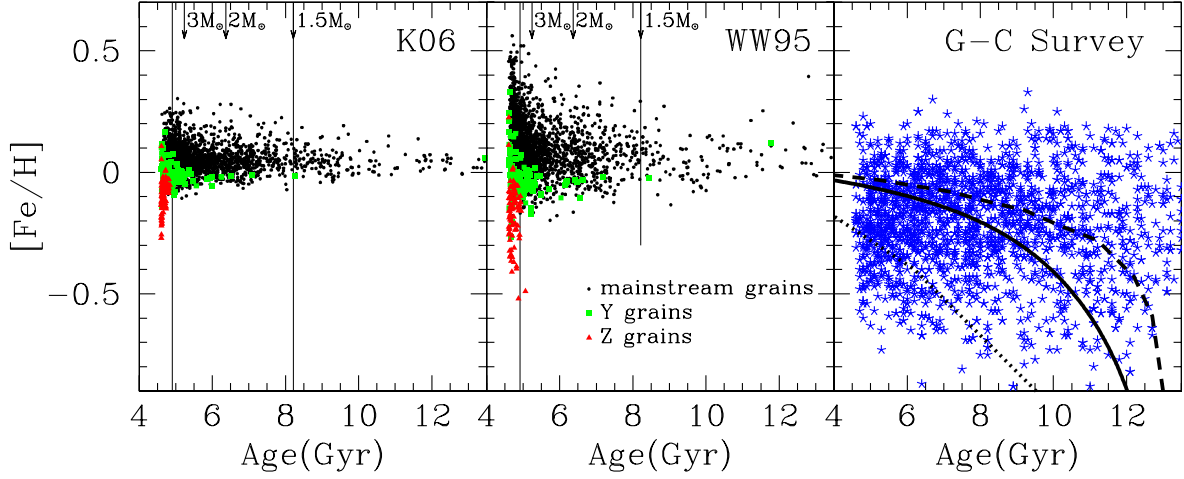}
\caption{
The age-metallicity relation 
(AMR) derived for the parent stars of SiC mainstream, Y, and Z grains using 
GEtool with two different sets of SNII yields: K06 \citep[][left
panel]{kobayashi06} and WW95 \citep[][middle panel]{woosley95}.
Note that the plotted ages are lower limits based on the experimental
2$\sigma$ error bar. The upper limits are undetermined 
(because of negative $\Delta^{30}$Si) and $>$13 Gyr for  
54\% and 26\% of the mainstream grains, respectively, while
they are $<$6 Gyr for 86\% of the Z grains.
Selected initial stellar masses are 
indicated at the top of the panels in correspondance to their ages.
Note that only AGB stars in the mass range between $\sim$1.5 \msun 
and $\sim$4 \msun are expected to become C rich and produce SiC grains 
\citep{groenewegen95,gail09}. The corresponding age limits 
are highlighted by the two vertical thin black lines, however, 
we did not remove 
the points outside this range because (i) they are lower limits and (ii) they 
could shift inside the allowed range when considering 
the several uncertainties discussed in the text.
The right panel shows the AMR obtained from the Geneva-Copenhagen (G-C) 
survey for 2037 stars 
in the 
solar neighbourhood with age uncertainties lower than 25\%, as compared
to the AMRs predicted by the different GCE models: 
GEtool (solid line, which is independent of the choice of the 
yields); TC96 \citep[dashed line,][]{timmes96}; 
KKU11 \citep[dotted line,][]{kobayashi11a}.
\label{fig:AMR}}
\end{figure}

Nonwithstanding the uncertanties discussed above, relatively,
the ages derived 
for the parent stars of the Y grains are similar to those of 
the mainstream grains, while those derived for the Z grains cover a 
much narrower range indicating that Z grains have on average parent 
stars of higher mass than mainstream and Y grains. This result 
supports proton captures at the base of the convective envelope (also 
known as ``hot bottom burning'') as the process responsible for 
lowering the $^{12}$C/$^{13}$C ratios in the Z grains (ranging from 20 
to 100) with respect to those observed in the Y grains ($>$ 100, by 
definition). If this interpretation is correct, GCE models are not 
required to match the $\delta^{29}$Si versus [Fe/H] relationship of 
\citet{zinner06} (Figure~\ref{fig:GCEmodels}), which was derived using 
AGB models of fixed mass lower than $\sim$ 3 M$_{\odot}$. On the other 
hand, one may wonder why there are no SiC grains from low-metallicity 
and low-mass parent stars. A possibility is that the very high C/O 
ratio reached in these stars (up to 20 - 30) may favour production of 
amorphous carbon dust rather than SiC \citep{sloan08}.

The SiC [Fe/H] distribution is determined by the steepness of 
the $\delta^{29}$Si versus [Fe/H] relationship, which varies with the 
choice of the core-collapse supernova (SNII) yields and the GCE model 
(Figure~\ref{fig:GCEmodels}). The {\it GEtool} \citep{fenner03a} chemical 
evolution model was computed with dual infall (i.e., a rapid formation 
of the halo, followed by subsequent, protracted, disk formation), 
initial mass function from \citet{kroupa93}, and a 
Schmidt-Kennicutt star formation prescription, and it is tuned to 
recover the gas and stellar abundances and radial surface densities in 
the Milky Way. In Figure~\ref{fig:AMR} we plot two SiC AMRs obtained using 
the most and the least steep $\delta^{29}$Si versus [Fe/H] relationships 
from the GEtool simulations. These corresponds to using the yields by 
\citet{woosley95} and \citet{kobayashi06}, respectively. 
The GEtool model computed using the 
SNII yields from \citet{kobayashi06} and that computed 
using \citet{chieffi04} produce a very similar SiC AMR 
with a [Fe/H] spread of a factor of three, within that observed 
in the G-C survey. \citet{timmes96} and \citet{kobayashi11a} obtained a much 
steeper and much flatter, respectively, $\delta^{29}$Si and [Fe/H] 
relationship than the GEtool models presented here. The 
difference depends on many of the ingredients of the GCE 
simulation, e.g., the initial mass function and the infall scheme, which 
also affect the theoretical AMR predicted directly by the GCE models. 
Interestingly, GCE models that predict a steeper $\delta^{29}$Si 
versus [Fe/H] relationship also predict a flatter AMR.
As a consistency check, in the right panel of Figure~\ref{fig:AMR} 
the AMRs predicted by the different GCE models 
considered here are compared to
the G-C stars. None of the models can recover the 
observed AMR. A better match may be found by recent models that
consider dynamics together with the chemical evolution in the Galaxy
\citep{kobayashi11b,pilkington12}. However, we do not consider them here 
as they have not been extended yet to 
include the evolution of isotopic abundances.
{\it Overall, the SiC grain data confirm the result of the G-C survey 
that stars exist with ages older than the Sun and metallicities 
higher than the Sun.} 

\begin{figure}
\includegraphics[angle=0,width=\textwidth]{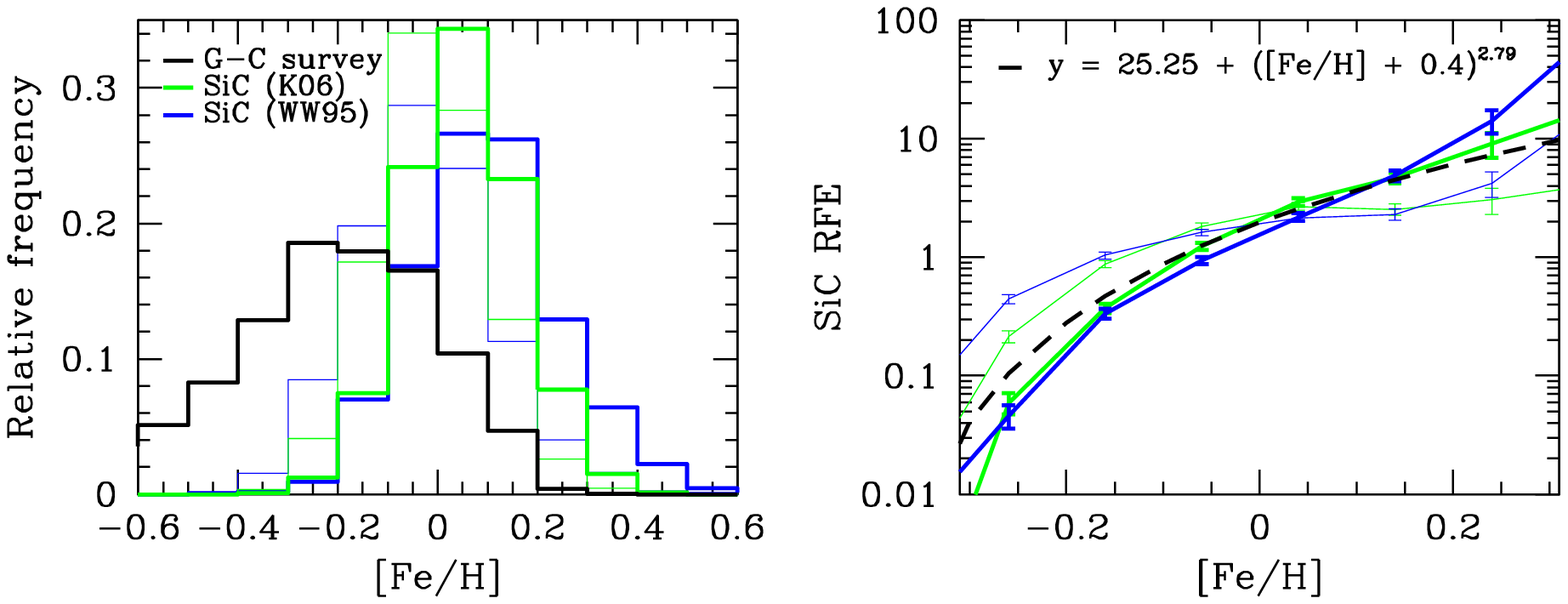}
\caption{The metallicity distribution
function (MDF) of the parent stars of 
mainstream SiC grains derived using GEtool and two different sets of SNII 
yields: WW95 \citep{woosley95} and K06 \citep{kobayashi06}, is compared to 
that obtained 
for stars in the solar neighbourhood from the Geneva-Copenhagen (G-C) 
survey (left panel).  
The SiC relative formation efficiency (RFE) 
of as function of the [Fe/H] is shown in 
the right panel. The thin lines present the results obtained 
renormalising $\delta^{29}$Si=50$\permil$ at [Fe/H]=0 and the black dashed line 
is a proposed power-law fit.
\label{fig:MDFdust}}
\end{figure}

To make a quantitative comparison of the metallicity distribution 
function (MDF) derived for the SiC grain parent stars and for the G-C stars 
we restrict ourselves to the mainstream SiC 
grains as they represent the least biased sample. We removed the Y and Z 
grains because their numbers are probably overestimated due to specific 
searches dedicated to identifying these types of grains. In the left 
panel of Figure~\ref{fig:MDFdust} we compare the MDF obtained from the SiC 
AMR to that obtained from the G-C survey.
Because the 1$\sigma$ error bar on the grain [Fe/H] 
that derives from the experimental uncertainty of $\delta^{29}$Si is
much smaller ($<$0.03 dex) than that of the stellar [Fe/H] ($\sim$0.1
dex), for better comparison we convolved the grain [Fe/H] data with a
Gaussian of $\sigma$=0.1.
The main difference between the grain and the stellar MDF
is that the mean metallicity of the parent stars of 
the mainstream SiC grains is $\simeq$ 50\% higher than that of the G-C 
stars. We interpret this as a selection effect between the grain and the 
star samples indicating that formation of SiC dust is favored by higher 
metallicity. We define the SiC relative formation efficiency (RFE) 
as the 
ratio between the normalised number of mainstream SiC grains and of G-C 
stars in each metallicity bin and plot these values in the right panel 
of Figure~\ref{fig:MDFdust}. The SiC RFE is unitless and its 
values are not absolute, but have a 
meaning only when considered relatively to each 
other.
We infer that the SiC RFE 
can be represented by a power law in metallicity. The relationship 
is well defined only for values of [Fe/H] between $\pm 0.3$ dex because 
in this range there are sufficient numbers of both mainstream SiC grains 
and G-C stars to make their ratio statistically meaningful.

An important systematic uncertainty in the derivation of the 
SiC RFE is related to the renormalised value of 
$\delta^{29}$Si=0 at [Fe/H]=0 (Item 1 of Sec.~\ref{sec:method}) 
Such renormalisation may easily 
be a few percent wrong, in particular due to the effect of 
inhomogeneities in the interstellar medium, which produce 
variations in $\delta^{29}$Si of the order of 70$\permil$ for the same 
[Fe/H] within $\pm0.01$ \citep{nittler05}. 
If $\delta^{29}$Si is renormalised to a 
positive value instead of zero, clearly the grain MDF becomes 
closer to that of 
the G-C survey and the derived SiC RFE is less 
steep. In Figure~\ref{fig:MDFdust} we also present the results 
obtained by setting $\delta^{29}$Si=50$\permil$ at [Fe/H]=0. 
The SiC RFE still 
increases with the metallicity, though the increase is less pronounced and 
disappears between 0$<$[Fe/H]$<$0.1. We consider this 
example as an upper limit: a choice of $\delta^{29}$Si$\sim$30$\permil$ at [Fe/H]=0
is probably more realistic being the value shown 
by the largest number of SiC, according to Figure 13 of \citet{nittler03}.

\section{Conclusions} \label{sec:conclusions}

The results presented here confirm that the relationship between age and 
metallicity in the Galaxy is relatively flat and that a spread of 
metallicities is present for each age. This cannot be recovered by 
traditional GCE models and requires a more sophisticated approach 
including coupling of Galactic dynamics and chemical evolution 
\citep{kobayashi11b,pilkington12}.  
Scattering and radial migration must have played an important 
role in determining the properties of stars in the  
solar neighbourhood \citep{clayton97}, as supported by 
recent observational studies \citep[e.g.][]{boeche13,ramirez13}.

We have interpreted the shift of the SiC MDF to higher metallicities than 
the G-C survey as a selection effect and derived that the SiC 
relative formation efficiency
increases with the stellar metallicity as a power law.
This result is in qualitative agreement with Spitzer observations of 
C-rich AGB stars in the Large and Small Magellanic Clouds 
\citep{sloan08}, which indicate that the mid-infrared emission from SiC 
and silicate dust decreases with the metallicity, while the 
emission from amorphous carbon dust does not. On the basis of this 
evidence, we tentatively predict that the MDF of stardust silicates 
\citep{mostefaoui04,nguyen10} should be similar to that of SiC grains, 
while the MDF of stardust graphite grains from C-rich AGB stars 
\citep{jadhav08} should be similar to that of the G-C survey. 
\citet{nittler09} conducted a similar exercise to ours based on the O 
ratios in stardust oxide grains and found evidence for the existence of 
a moderate AMR. If the $^{18}$O/$^{16}$O ratio is a good 
indicator of the stellar metallicity, we expect  
a correlation between $^{18}$O/$^{16}$O and $\delta^{29}$Si, which 
is not shown by the stardust 
silicate grains, but could be masked by
the effect of dilution with normal material \citep{nguyen10}. 
It should also be 
noted that stardust oxide and silicate grains are more likely to come 
from lower-mass stars than SiC grains \citep{gail09}, which may result 
in a different AMR. It will be possible to statistically investigate 
these issues in the future when more high-precision Si data for these 
types of stardust are available.

\acknowledgments

The first two authors have equally contributed to this paper.
We thank Chiaki Kobayashi and Mike Savina for 
discussion and Keith Hsuan for technical support. We thank
the anonymous referee for helping us to greatly improve 
the treatment and discussion of the uncertainties.
ML is an ARC Future Fellow and Monash Fellow. This project was supported by 
the Monash University ``Advancing Women in Research'' grant. 


\clearpage




\clearpage

\end{document}